\documentstyle[12pt,amssymbols]{article}

\def\journal{\topmargin .3in        \oddsidemargin .5in
        \headheight 0pt        \headsep 0pt
        \textwidth 5.625in 
        \textheight 8.25in 
        \marginparwidth 1.5in
        \parindent 2em
        \parskip .5ex plus .1ex                \jot = 1.5ex}
\journal

\catcode`\@=11
\def\marginnote#1{}
\def\titlepage{\@restonecolfalse\if@twocolumn\@restonecoltrue\onecolumn
     \else \newpage \fi \thispagestyle{empty}\c@page\z@
        \def\thefootnote{\fnsymbol{footnote}} }
\def\endtitlepage{\if@restonecol\twocolumn \else \newpage \fi
        \def\thefootnote{\arabic{footnote}}
        \setcounter{footnote}{0}}  
\catcode`@=12
\relax
\def\reflist{\section*{References\markboth
        {REFLIST}{REFLIST}}\list
        {[\arabic{enumi}]\hfill}{\settowidth\labelwidth{[999]}
        \leftmargin\labelwidth
        \advance\leftmargin\labelsep\usecounter{enumi}}}
 \relax
\catcode`\@=11
\def\section{\@startsection {section}{1}{0pt}{-3.5ex plus -1ex minus
 -.2ex}{2.3ex plus .2ex}{\raggedright\large\bf}}
\catcode`\@=12
\newskip\humongous \humongous=0pt plus 1000pt minus 1000pt

\newif\ifdtup

\def\oldreffmt#1{\rlap{[#1]} \hbox to 2\parindent{}}

\def\figfmt#1{\rlap{Figure {#1}} \hbox to 1in{}}

\def\beq{\begin{equation}}
\def\eeq{\end{equation}}

\def\bea{\begin{eqnarray}}

\relax

\def\thefootnote{\fnsymbol{footnote}}

\def \gothic {\Bbb}
\def \lteq {\le}
\def \mod {\; \hbox{mod} \;}

\def\blank#1#2#3{#1 (19#3) #2}
\def\genjour#1#2#3#4{#1\ \blank{#2}{#3}{#4}}

\def\cpc#1#2#3{ \genjour {Comput.\ Phys.\ Commun.}       {#1} {#2} {#3}}
\def\jp#1#2#3{  \genjour {J.\ Phys.}                     {#1} {#2} {#3}}
\def\jmp#1#2#3{ \genjour {J.\ Math.\ Phys.}              {#1} {#2} {#3}}
\def\np#1#2#3{  \genjour {Nucl.\ Phys.}                  {#1} {#2} {#3}}
\def\pl#1#2#3{  \genjour {Phys.\ Lett.}                  {#1} {#2} {#3}}

\def\prep#1#2#3{\genjour {Phys.\ Rep.}                   {#1} {#2} {#3}}

\def\rmp#1#2#3{ \genjour {Rev.\ Mod.\ Phys.}             {#1} {#2} {#3}}

\def \mult{\otimes}

\def \bar{\overline}
\def \space{\vskip .25in}

\def \f#1#2{$\frac{#1}{#2}$}

\def \shw {\hbox{shw}}

\newfont{\bff}{cmtcsc10 scaled \magstep 1}

\def \samplecaption#1#2{\begin{center}
                        \begin{tabular}{p{1in}p{4.2in}}
                        Sample #1:  & #2
                        \end{tabular}
                        \end{center}
                        \vskip .2in}

\def \explanatory#1#2{\noindent
                      \begin{tabular}{p{1.7in}p{3.7in}}
                      \noindent {\tt #1} & #2
                      \end{tabular}
                      \par}

\begin{document}

\begin{titlepage}
\begin{center}
November 20, 1995
           \hfill    \begin{tabular}{l}
                         LBL-37969 \\
                     \end{tabular}

\vskip .5in

{\large \bf Program for generating tables of $SU(3)$ coupling
coefficients}
\vskip .25in

Thomas A.\ Kaeding\footnote{This
work was supported by the Director, Office of Energy Research, Office of
High Energy and Nuclear Physics, Division of High Energy Physics of the
U.S. Department of Energy under Contract
DE-AC03-76SF00098.}\footnote{Electronic address:
  {\tt takaeding@lbl.gov}.}

{\em Theoretical Physics Group,
     Lawrence Berkeley Laboratory\\
     University of California,
     Berkeley, California  94720}

\vskip .15in
H.\ Thomas Williams\footnote{Electronic address:
  {\tt twilliam.ht@wlu.edu}.}

{\em Department of Physics,
     Washington and Lee University\\
     Lexington, Virginia  24450}

\end{center}

\vskip .25in

\begin{abstract}
\noindent
A C-Language program which tabulates the isoscalar factors and
Clebsch-Gordan coefficients for products of representations in
$SU(3)$ is presented.
These are efficiently computed using recursion relations, and the
results are presented in exact precision as square roots of rational
numbers.
Output is in \LaTeX\ format.
\end{abstract}

\vskip 10pt \noindent
\begin{tabular}{p{1in}p{4in}}
{\it Keywords:} & $SU(3)$, Clebsch-Gordan coefficients, Wigner
                  coefficients,
                  vector coupling coefficients, isoscalar factors
\end{tabular}

\vskip 10pt \noindent
\begin{tabular}{p{1in}p{4in}}
{\it Classification:} &
          4.2  Computational methods of algebras and groups
\end{tabular}

\begin{center}
\vskip 10pt
Submitted to {\it Computer Physics Communications}
\end{center}

\end{titlepage}

\renewcommand{\thepage}{\roman{page}}
\setcounter{page}{2}
\mbox{ }

\vskip 1in

\begin{center}
{\bf Disclaimer}
\end{center}

\vskip .2in

\begin{scriptsize}
\begin{quotation}

This document was prepared as an account of work sponsored by the United
States Government. While this document is believed to contain correct
information, neither the United States Government nor any agency
thereof, nor The Regents of the University of California, nor any of
their employees, makes any warranty, express or implied, or assumes
any legal liability or responsibility for the accuracy, completeness,
or usefulness of any information, apparatus, product, or process
disclosed, or represents that its use would not infringe privately
owned rights.  Reference herein to any specific commercial products
process, or service by its trade name, trademark, manufacturer, or
otherwise, does not necessarily constitute or
imply its endorsement, recommendation, or favoring by the United States
Government or any agency thereof, or The Regents of the University of
California.  The views and opinions of authors expressed herein do not
necessarily state or reflect those of the United States Government or
any agency thereof, or The Regents of the University of California.

\end{quotation}
\end{scriptsize}

\vskip 2in

\begin{center}
\begin{small}
{\it Lawrence Berkeley Laboratory is an equal opportunity employer.}
\end{small}
\end{center}

\newpage
\renewcommand{\thepage}{\arabic{page}}
\setcounter{page}{1}

\begin{center}
{\large \bf PROGRAM SUMMARY}
\end{center}

\vskip 10pt \noindent
{\it Title of program:}  SU3

\vskip 10pt \noindent
{\it Catalogue number:}

\vskip 10pt \noindent
{\it Program obtainable from:}  CPC Program Library, Queen's University
of Belfast, N.\ Ireland (see application form in this issue)

\vskip 10pt \noindent
{\it Licensing provisions:}  Persons requesting the program must sign
the standard CPC non-profit use license (see license agreement in
every issue).

\vskip 10pt \noindent
{\it Computers for which the program is designed and tested:}
DEC VAX 4000/90, DEC Alpha 3000/700, Dell 486, Dell Pentium, HP715/33

\vskip 10pt \noindent
{\it Operating systems under which the program has been tested:}
VMS v6.2, MSDOS6.x, HP UNIX 9.0

\vskip 10pt \noindent
{\it Programming language used:}  C Language

\vskip 10pt \noindent
{\it Memory required to execute:}
Variable;  on the VAX it requires
934 k words for {\bf 3} $\otimes$ {\bf 3},
998 k words for {\bf 27} $\otimes$ {\bf 27}

\vskip 10pt \noindent
{\it Disk space used for output:}
Variable;  2 k words for {\bf 3} $\otimes$ {\bf 3},
182 k words for {\bf 27} $\otimes$ {\bf 27}

\vskip 10pt \noindent
{\it No.\ of bits in a word:}  32

\vskip 10pt \noindent
{\it No.\ of processors used:}  1

\vskip 10pt \noindent
{\it Has the code been vectorized?}  No

\vskip 10pt \noindent
{\it No.\ of lines in distributed program:}  2601


\vskip 10pt \noindent
{\it Classification:}  4.2  Computational methods of algebras and groups

\vskip 10pt \noindent
{\it Nature of physical problem:}
Calculations using models based on $SU(3)$ symmetry often require the
Clebsch-Gordan coefficients which give the amplitudes for the expansion
of direct products of irreducible representations.

\vskip 10pt \noindent
{\it Method of solution:}
Recursion relations [1]
are used to construct tables of the isoscalar factors for
the product of two irreducible representations. The $SU(2)$
Clebsch-Gordan coefficients
[2] [3]
are then used with the isoscalar factors to
produce the $SU(3)$ Clebsch-Gordan coefficients.

\vskip 10pt \noindent
{\it Restrictions on the complexity of the problem:}
The program is limited by computer memory and maximum size of the
integer variables.  On the DEC VAX with 32-bit integers, the largest
product that has been successfully computed is
{\bf 27} $\otimes$ {\bf 27}.

\vskip 10pt \noindent
{\it Typical running time:}
Running time varies according to the sizes of $p_1$, $q_1$, $p_2$,
and $q_2$ in the product $(p_1, q_1) \otimes (p_2, q_2)$.
On the VAX for {\bf 3} $\otimes$ {\bf 3} the CPU time is 1.2 sec;
for {\bf 27} $\otimes$ {\bf 27} the CPU time is 13 min 11 sec.
On the Alpha, the latter time is only 1 min 13 sec.

\vskip 10pt \noindent
{\it Unusual features of the program:}
The program has these unique features:  it determines the
representations in the Clebsch-Gordan series;  it uses recursion
relations to efficiently compute the isoscalar factors;  it presents
{\it exact} results; it provides results as \LaTeX\ [4] formatted
tables.

\vskip 10pt \noindent
{\it References}

\noindent
\begin{tabular}{lp{5.24in}}
[1] & H.\ T.\ Williams, Symmetry Properties of SU3
      Vector Coupling Coefficients, WLUPY-9-93.\\
\end{tabular}

\noindent
\begin{tabular}{lp{5.24in}}
[2] & E.\ U.\ Condon and G.\ H.\ Shortley, The Theory of Atomic Spectra
      (Cambridge University Press, London, 1957).\\
\end{tabular}

\noindent
\begin{tabular}{lp{5.24in}}
[3] & E.\ P.\ Wigner, Group Theory and its Application to
      the Quantum Mechanics of Atomic Spectra,
      trans.\ by J.\ J.\ Griffin
      (Academic Press, New York, 1959).\\
\end{tabular}

\noindent
\begin{tabular}{lp{5.24in}}
[4] &  L.\ Lamport, \LaTeX:  A Document Preparation System
      (Addison-Wesley, Menlo Park CA, 1986).\\
\end{tabular}

\newpage

\begin{center}
{\large \bf LONG WRITE-UP}
\end{center}

\section{Introduction}

In calculations in nuclear and particle physics which assume $SU(3)$
symmetry, it is often useful to have values for relevant Clebsch-Gordan
coefficients (CGC) \footnote{The Clebsch-Gordan coefficients are also
called {\it vector coupling
coefficients} or {\it Wigner coefficients}.  Readers interested in a
good theoretical background are directed to \cite{deswart}
\cite{carruthers} \cite{georgi} \cite{bied}.}.
Some tables of limited extent have been
published in the past \cite{deswart} \cite{chilton} \cite{wali}
\cite{hecht} \cite{rashid} \cite{kaeding2}. Also, programs have been
written which generate these coefficients \cite{akiyama}
\cite{williams1} \cite{williams2} \cite{kaeding1}. This new program
is meant to replace them.
It has some advantages that make it faster and easier to use.
Comparative speed is achieved by use of recursion relations to
generate the isoscalar factors (ISF), from which the CGC are
constructed.  This method requires
less memory and allows larger tables to be
generated.  All calculations are done exactly, using integer variables.
Output is in the form of \LaTeX\ \cite{latexref} formatted tables.
The code is written in ANSI standard C, so it should run on most
conventional platforms.

\section{Preliminaries}

The Clebsch-Gordan coefficients are the amplitudes for the projection
of the product of two irreducible representations (${\bf r}_1$,
${\bf r}_2$) of  $SU(3)$ onto the separate irreducible representations
(${\bf R}_n$) found in the Clebsch-Gordan series:
  \begin{equation}
    {\bf r}_1 \otimes {\bf r}_2 = \sum {\bf R}_n.
  \label{eq:series}
  \end{equation}
They can be defined by
  \begin{equation}
    \left| {\bf r}_1 \alpha_1 \right> \left| {\bf r}_2 \alpha_2 \right>
        = \sum
        \left< {\bf R}_i A_i |
            {\bf r}_1 \alpha_1 {\bf r}_2 \alpha_2 \right>
        \left| {\bf R}_i A_i \right>,
  \end{equation}
where $A_i$, $\alpha_1$, and $\alpha_2$ denote the quantum numbers
specifying the particular states within the representations.
Particle physicists are accustomed to the quantum numbers hypercharge
$y$, isospin $i$, and third component of isospin $i_z$;  herein are
also used the ``projection'' quantum numbers $k$, $l$, $m$, which are
related to $y$, $i$, and $i_z$ in a representation (denoted by the
representation variables $(p, q)$) by
  \begin{equation}
  \begin{array}{ccl}
    k &=& \frac {1}{3} (p + 2q) + \frac{1}{2} y + i, \\
    l &=& \frac {1}{3} (p + 2q) + \frac{1}{2} y - i, \\
    m &=& \frac {1}{3} (p + 2q) + \frac{1}{2} y + i_z, \\
  \end{array}
  \end{equation}
The highest-weight state (shw) in a given representation is defined to
be that which has the largest $i_z$;  thus the state with
  \begin{equation}
  \begin{array}{c}
    k_{\hbox{\scriptsize shw}} = m_{\hbox{\scriptsize shw}} = p + q, \\
    l_{\hbox{\scriptsize shw}} = 0. \\
  \end{array}
  \end{equation}

The (outer) degeneracy of an irreducible representation in the
Clebsch-Gordan series (Equation \ref{eq:series}) can be determined
algebraically from the representation variables \cite{oreilly}.
A representation appears in the series if the degeneracy is greater or
equal to one, and only in such cases does the program attempt to
calculate its coupling coefficients.  If we represent the irreducible
representations as
  \begin{equation}
  \nonumber
  \begin{array}{lcl}
    {\bf r}_1 &=& (p_1, q_1), \\
    {\bf r}_2 &=& (p_2, q_2), \\
    {\bf R} &=& (p, q);
  \end{array}
  \end{equation}
and define
  \begin{equation}
  \nonumber
  \begin{array}{lcl}
    a &=& \frac{1}{3}
          \left[ (p_1 - q_1) + (p_2 - q_2) - (p - q) \right], \\
    b &=& \frac{1}{3}
          \left[ (p_1 + 2q_1) + (p_2 + 2q_2) - (p + 2q) \right];
  \end{array}
  \end{equation}
then if both
 \begin{enumerate}
  \item  $a  \in {\gothic Z}$ (integers) (thus $b \in {\gothic Z}$), and
  \item  $0 \lteq b \lteq \hbox{min} (q_1 + q_2, p_1 + q_1, p_2 + q_2)$,
         $-\hbox{min} (q_1, q_2) \lteq a \lteq \hbox{min} (p_1, p_2)$,
         $0 \lteq
          a + b \lteq \hbox{min} (p_1 + q_1, p_2 + q_2, p_1 + p_2)$,
 \end{enumerate}
the representation appears in the series and has degeneracy
  \begin{equation}
  \begin{array}{lll}
    d (p, q: p_1, q_1; p_2, q_2)
      & = & 1 \\
      & + & \hbox{min} (q_2, p_1 + q_1, b, p_1 - a) \\
      & - & \hbox{max} (0, b - q_1, b - p_2, -a, b - a - q_1,
                              a + b - p_2).
  \end{array}
  \end{equation}

The isoscalar factors $F$ are defined by
  \begin{equation}
  \begin{array}{lcl}
    \left< {\bf R} \; Y \; I \; I_z | {\bf r}_1 \; y_1 \; i_1 \; i_{1z}
                        \; {\bf r}_2 \; y_2 \; i_2 \; i_{2z} \right>
    &=& F ({\bf R}, Y, I: {\bf r}_1, y_1, i_1; {\bf r}_2, y_2, i_2) \\
    & & \times \left< I \; I_z | i_1 \; i_{1z} \; i_2 \; i_{2z} \right>,
  \end{array}
  \label{eq:definitionisf}
  \end{equation}
where the last bracket is an $SU(2)$ CGC,
which in the Condon-Shortley phase convention is
\cite{wigner2} \cite{condon}
  \begin{eqnarray}
    && \left< I \; I_z | i_1 \; i_{1z} \; i_2 \; i_{2z} \right> \quad =
                 \nonumber \\
    && \quad \quad \delta (I_z - i_{1z} - i_{2z}) \nonumber \\
    && \quad \quad \times
        \sqrt{\frac{(I+i_1-i_2)!(I-i_1+i_2)!(i_1+i_2-I)!(I+I_z)!(2I+1)}
               {(I+i_1+i_2+1)!(i_1-i_{1z})!
                (i_1+i_{1z})!(i_2-i_{2z})!(i_2+i_{2z})!}} \\
    && \quad \quad \times
        \sum_n \frac{(-1)^{n+i_2+i_{2z}}
                      (I+i_2+i_{1z}-n)!(i_1-i_{1z}+n)!}
                     {(I-i_1+i_2-n)!(I+I_z-n)!n!(i_1-i_2-I_z+n)!},
                \nonumber
  \label{eq:su2clebsch}
  \end{eqnarray}
where $n$ runs from
  \begin{equation}
  \nonumber
    n_{\hbox{\scriptsize low}} = \hbox{max} (0, i_{1z}-i_1, I_z-i_1-i_2)
  \end{equation}
to
  \begin{equation}
  \nonumber
    n_{\hbox{\scriptsize high}}
        = \hbox{min} (I+i_2+i_{1z}, I-i_1+i_2, I+I_z).
  \end{equation}
Note that
  \begin{equation}
    \left< I \; I_z | i_1 \; i_{1z} \; i_2 \; i_{2z} \right>
        = (-1)^{I - i_1 - i_2}
        \left< I \; I_z | i_2 \; i_{2z} \; i_1 \; i_{1z} \right>.
  \label{switchisos}
  \end{equation}

\subsection{Conventions}

The internal phase convention, fixing totally the relative phases
between states within a representation, is chosen to be that of de Swart
\cite{deswart}.  The irreducible representations of $SU(3)$ can be
thought of as consisting of $SU(2)$ multiplets ({\it isomultiplets}),
each at a specific hypercharge.  This internal
phase convention corresponds to adopting the Condon-Shortley
\cite{condon} phase convention for the $T$- and $V$-spin operators.
(In the flavor-$SU(3)$ notation, $T^{\pm}$ interchanges $u$ and
$d$ quarks, while $V^{\pm}$ interchanges $u$ and $s$.)

The overall phase of the representations in the product is also
chosen following de Swart \cite{deswart}.  Consider the state
of highest weight in the product representation,
$ \left| {\bf R} \; \hbox{shw} \right> $;
the state of highest weight in the first factor,
$ \left| {\bf r}_1 \; \hbox{shw}_1 \right> $;
and the state in the second factor {\bf r}$_2$ = $(p_2, q_2)$
with the highest isospin that couples to shw and shw$_1$,
$ \left| {\bf r}_2 \; k_{2 \hbox{\scriptsize max}}(\hbox{shw},
\hbox{shw}_1) \; l_{2 \hbox{\scriptsize min}}(\hbox{shw},
\hbox{shw}_1) \right> $.
The outer phase conventions requires that the Clebsch-Gordan coefficient
for coupling these states be real and positive:
  \begin{equation}
    \left< {\bf R} \; \hbox{shw} | {\bf r}_1 \; \hbox{shw}_1 \;
           {\bf r}_2 \;
           k_{2 \hbox{\scriptsize max}}(\hbox{shw}, \hbox{shw}_1) \;
           l_{2 \hbox{\scriptsize min}}(\hbox{shw}, \hbox{shw}_1)
    \right> \; > \; 0.
  \end{equation}
The Condon-Shortley convention for the $SU2$ CGC assures that the
corresponding isoscalar factor is also real and positive.  The internal
and external phase conventions guarantee that all isoscalar factors and
Clebsch-Gordan coefficients are real.

In couplings where the degeneracy $d > 1$ there are $d$ distinct sets
of CGC, and one must chose by convention a technique for resolving the
outer  degeneracy.  Several distinct techniques for this resolution are
described in the literature \cite{alcaras} \cite{bied2}.
The technique chosen here considerably simplifies the determination of
isoscalar factors through use of recursion relations, and produces ISF
and CGC which share the symmetries under interchange of representations
(Racah symmetries), which are familiar from the $SU2$ case.  The
technique relies on the fact that there are nonvanishing ISFs of the
form
  \begin{equation}
  \nonumber
    F ({\bf R}, \hbox{shw}: {\bf r}_1, \hbox{shw}_1;
                            {\bf r}_2, k_2, l_2)
  \end{equation}
for at least as many distinct values of $k_2$ as the degeneracy $d$.
If one chooses ISFs for the $d-1$ highest values of $k_2$ to be zero,
the recursion relations, normalization, and the above-mentioned sign
conventions uniquely determine a full set of
ISFs.  A second set can be determined by choosing the ISFs for the
$d-2$ highest values of $k_2$ to be zero, and insisting that this
second set be orthogonal to the first.  Subsequent sets of ISFs are
formed likewise by successively forcing fewer of these ISFs to zero,
and enforcing the condition of orthogonality to
those previously determined.

This technique of resolution of the outer degeneracies insures that
the coupling coefficients in all cases change by at most a sign under
the interchange of {\bf r}$_1$ and {\bf r}$_2$, in contrast to the
``canonical'' resolution scheme \cite{bied2}.

\subsection{Symmetry Phases}

With the conventions of the previous section, all the ISFs and CGCs
for $SU(3)$ are fully determined.  The resulting symmetries are useful
in reducing the number of values which must necessarily be tabulated.
Three phases are involved in the symmetries and are defined in
\cite{deswart}.

In the case of interchange of the two factor representations,
  \begin{equation}
  \begin{array}{lcl}
    F({\bf R}, Y, I: {\bf r}_2, y_2, i_2; {\bf r}_1, y_1, i_1)
     & = & (-1)^{I - i_1 - i_2} \xi_1 ({\bf R}: {\bf r}_1; {\bf r}_2)\\
     &   & F({\bf R}, Y, I: {\bf r}_1, y_1, i_1; {\bf r}_2, y_2, i_2),
  \end{array}
  \label{defxi}
  \end{equation}
where the factor $(-1)^{I - i_1 - i_2}$ comes from Equation
\ref{switchisos}.
The phase $\xi_1 ({\bf R}: {\bf r}_1; {\bf r}_2)$
depends only on the identity of the representations
and on our phase conventions.
Consider the reversed product ${\bf r}_2$ $\mult$ {\bf r}$_1$.
Suppose that the highest-isospin multiplet in {\bf r}$_1$ that
couples to the state of highest weight in {\bf r}$_2$
and the shw of {\bf R} with nonzero ISF has quantum numbers
$y_1'$ and $i_1'$.
Then
  \begin{equation}
    \xi_1 ({\bf R}: {\bf r}_1; {\bf r}_2)
        = (-1)^{I_{\hbox{\scriptsize h}} -
                i_1' - i_{2 \hbox{\scriptsize h}}}
        \times \hbox{sign} \left[ F ({\bf R}, \hbox{shw}:
            {\bf r}_1 , y_1', i_1';
        {\bf r}_2, \hbox{shw}_2) \right],
  \end{equation}
where $\hbox{sign}(x)$ = $x / |x|$.

Should the second factor and the product representations be
interchanged, a
multiplicative phase and a term related to the ratio of
the representations' dimensions is generated:
  \begin{eqnarray}
    && F(\bar{\bf r}_2, \hbox{-}y_2, i_2: {\bf r}_1, y_1, i_1;
           \bar{\bf R}, \hbox{-}Y, I) \quad = \nonumber \\
    && \quad \quad (-1)^{i_1+y_1/2}
                    \xi_2 ({\bf R}: {\bf r}_1; {\bf r}_2)
             \nonumber \\
    && \quad \quad \times  \sqrt{\frac{(p_2+1)(q_2+1)(p_2+q_2+2)(2I+1)}
                    {(p+1)(q+1)(p+q+2)(2i_2+1)}} \label{defxi2} \\
    && \quad \quad \times F({\bf R}, Y, I: {\bf r}_1, y_1, i_1;
                                   {\bf r}_2, y_2, i_2).
            \nonumber
  \end{eqnarray}

If all the representations are conjugated, then a phase $\xi_3$ enters:
  \begin{equation}
  \begin{array}{lcl}
    F(\bar{\bf R}, Y, I: \bar{\bf r}_1, y_1, i_1;
           \bar{\bf r}_2, y_2, i_2)
       & = & (-1)^{I-i_1-i_2} \xi_3 ({\bf R}: {\bf r}_1; {\bf r}_2) \\
       &   & \times
               F({\bf R}, \hbox{-}Y, I: {\bf r}_1, \hbox{-}y_1, i_1;
                                   {\bf r}_2, \hbox{-}y_2, i_2).
  \end{array}
  \label{defzeta}
  \end{equation}
Consider the isomultiplets described in Section 2.1.
Then from Equation \ref{switchisos} we find that
  \begin{equation}
    \xi_3 ({\bf R}: {\bf r}_1; {\bf r}_2)
        = (-1)^{I_{\hbox{\scriptsize h}}
            - i_{1 \hbox{\scriptsize h}} - i_2'}.
  \end{equation}

\section{Calculational Method}

\subsection{Recursion Relations for Isoscalar Factors}

Recursion relations
among the CGC for $SU(3)$ can be generated by
applying one of the $SU(3)$ ladder operators to a pair of $SU(3)$
single particle states coupled (via CGCs) to a state of good $SU(3)$
quantum numbers.  Using the linearity of the ladder operators
\cite{deswart} (e.g. $K_+ = K_{1+}+K_{2+}$,  where $K_+$ is the
ladder operator for the coupled state, and $K_{1+}$ and $K_{2+}$
are the corresponding operators for the 1 and 2 states), and
orthogonality of states with different $SU(3)$ quantum
numbers, one can find four-term recursion relations for the CGCs.
Making use of explicit analytic forms for the SU2 Clebsch-Gordan
coefficients these can be turned into four-term recursion relations
for the $SU(3)$ isoscalar factors.
Two of the ladder
operators ($I_+$ and $I_-$) generate recursions only within the SU2
variables and produce only identities for the ISF.  From the remaining
four ladder operators expressions fully sufficient to generate all the
$SU(3)$ ISFs have been derived.

The simplest language for presenting the recursion relations is to use
$(p,q)$ to indicate an irreducible representation, and $(k,l,m)$ to
specify a state within the representation.  The scheme involves, as a
first step,  using one of the following relations to establish values
for all the isoscalar factors representing coupling to the state of
highest weight (i.e. $k=m=p+q, l=0$).  A simplified notation which
omits the $p$'s and $q$'s, and which uses the symbol shw to
represent $k = p+q$, $l=0$ will be adopted.  Thus, for example,
  \begin{equation}
  \nonumber
    F(p,q,k=p+q,l=0:p_1,q_1,k_1,l_1;p_2,q_2,k_2,l_2)
         = F(\hbox{shw}:k_1,l_1;k_2,l_2).
  \end{equation}

The two recursion relations involving the states of highest weight are
  \begin{equation}
  \begin{array}{lcl}
    0 & = &   a_1 F(\shw:k_1-1,l_1;k_2,l_2)
            + a_2 F(\shw:k_1,l_1;k_2-1,l_2) \\
      &   & + a_3 F(\shw:k_1,l_1-1;k_2,l_2)
            + a_4 F(\shw:k_1,l_1;k_2,l_2-1).
  \end{array}
  \label{eq:a}
  \end{equation}
where
  \begin{eqnarray}
    a_1 & = & \sqrt{(k_1+1)(k_1-q_1)(p_1+q_1-k_1+1)
                    (p+q+k_1-l_1+k_2-l_2+3)} \nonumber \\
        &   & \times
              \sqrt{\frac{2
              (p+q+k_1-l_1-k_2+l_2+1)}{(k_1-l_1)(k_1-l_1+1)}},
                                                          \nonumber \\
    a_2 & = & a_1 (1 \leftrightarrow 2), \nonumber \\
    a_3 & = & -\sqrt{l_1(q_1-l_1+1)(p_1+q_1-l_1+2)} \\
        &   &  \times \sqrt{(-p-q+k_1-l_1+k_2-l_2+1)} \nonumber \\
        &   & \times
              \sqrt{\frac{2
              (p+q-k_1+l_1+k_2-l_2+1)}{(k_1-l_1+1)(k_1-l_1+2)}},
                                                         \nonumber \\
    a_4 & = & - a_3 (1 \leftrightarrow 2); \nonumber
  \end{eqnarray}
and
  \begin{equation}
  \begin{array}{lcl}
    0 & = &   b_1 F(\shw:k_1+1,l_1;k_2,l_2)
            + b_2 F(\shw:k_1,l_1;k_2+1,l_2) \\
      &   & + b_3 F(\shw:k_1,l_1+1;k_2,l_2)
            + b_4 F(\shw:k_1,l_1;k_2,l_2+1).
  \end{array}
  \end{equation}
where
  \begin{eqnarray}
    b_1 & = & \sqrt{(k_1+2)(k_1-q_1+1)(p_1+q_1-k_1)
                    (-p-q+k_1-l_1+k_2-l_2+1)} \nonumber \\
        &   & \times
              \sqrt{\frac{2
              (p+q-k_1+l_1+k_2-l_2+1)}{(k_1-l_1+1)(k_1-l_1+2)}},
                                                    \nonumber \\
    b_2 & = & - b_1 (1 \leftrightarrow 2), \nonumber \\
    b_3 & = & \sqrt{(l_1+1)(q_1-l_1)(p_1+q_1-l_1+1)
                    (p+q+k_1-l_1+k_2-l_2+3)} \\
        &   & \times
              \sqrt{\frac{2
              (p+q+k_2-l_2-k_2+l_2+1)}{(k_1-l_1)(k_1-l_1+1)}},
                                                    \nonumber \\
    b_4 & = & b_3 (1 \leftrightarrow 2). \nonumber
  \end{eqnarray}

Once the states of highest weight have been determined, two other
relations are sufficient to step from these to any non-shw state.
They are
  \begin{equation}
  \begin{array}{r}
    F(k,l:k_1,l_1;k_2,l_2)  =
                 c_1 F(k+1,l-1:k_1,l_1;k_2,l_2) \\
               + c_2 F(k,l-1:k_1,l_1-1;k_2,l_2) \\
               + c_3 F(k,l-1:k_1,l_1;k_2-1,l_2) \\
               + c_4 F(k,l-1:k_1,l_1;k_2,l_2-1),
  \end{array}
  \end{equation}
where
  \begin{eqnarray}
    c_1    & = & \alpha
                 \sqrt{\frac{(k+2)(k-q+1)(p+q-k)(k_1-l_1+k_2-l_2-k+l)}
                 {(k-l+2)^2(k-l+k_1-l_1+k_2-l_2+4)}}, \nonumber \\
    c_2    & = & 2 \alpha
                 \sqrt{\frac{l_1(q_1-l_1+1)(p_1+q_1-l_1+2)}
                 {(k_1-l_1+2)(k-l+k_1-l_1+k_2-l_2+4)}}, \nonumber \\
           &   & \times
                 \sqrt{\frac{(k_1-l_1+1)}{(k-l-k_1+l_1+k_2-l_2+2)}}
                                                      \nonumber \\
    c_3    & = & -\alpha
                 \sqrt{\frac{(k_2+1)(k_2-q_2)
                              (k_1-l_1+k_2-l_2-k+l)}
                 {(k_2-l_2)(k_2-l_2+1)}} \\
           &   & \times
                 \sqrt{\frac{(p_2+q_2-k_2+1)}
                            {(k-l-k_1+l_1+k_2-l_2+2)}},
                                                 \nonumber \\
    c_4    & = & \alpha
                 \sqrt{\frac{l_2(q_2-l_2+1)
                             (k-l+k_1-l_1-k_2+l_2+2)}
                 {(k_2-l_2+1)(k_2-l_2+2)}}
                                                 \nonumber \\
           &   & \times
                 \sqrt{\frac{(p_2+q_2-l_2+2)}
                            {(k-l+k_1-l_1+k_2-l_2+4)}};
                                                 \nonumber \\
  \end{eqnarray}
and
  \begin{equation}
    \alpha = \sqrt{\frac{(k-l+2)^2(k-l-k_1+l_1+k_2-l_2+2)}
                        {l(q-l+1)(p+q-l+2)(k-l+k_1-l_1-k_2+l_2+2)}};
  \end{equation}
and
  \begin{equation}
  \begin{array}{r}
    F(k,0:k_1,l_1;k_2,l_2)  =
                d_1 F(k+1,0:k_1+1,l_1;k_2,l_2) \\
              + d_2 F(k+1,0:k_1,l_1;k_2+1,l_2) \\
              + d_3 F(k+1,0:k_1,l_1;k_2,l_2+1),
  \end{array}
  \end{equation}
where
  \begin{eqnarray}
    d_1    & = & 2 \beta
                 \sqrt{\frac{(k_1+2)(k_1-q_1+1)}
                 {(k_1-l_1+2)(k-l+k_1-l_1+k_2-l_2+4)}}, \nonumber \\
           &   & \times
                 \sqrt{\frac{(p_1+q_1-k_1)(k_1-l_1+1)}
                            {(k-l+k_1-l_1-k_2+l_2+2)}} \nonumber \\
    d_2    & = & \beta
                 \sqrt{\frac{(k_2+2)(k_2-q_2+1)(p_2+q_2-k_2)}
                 {(k_2-l_2+1)(k_2-l_2+2)}} \\
           &   & \times
                 \sqrt{\frac{(k-l-k_1+l_1+k_2-l_2+2)}
                            {(k-l+k_1-l_1+k_2-l_2+4)}}, \nonumber \\
    d_3    & = & \beta
                 \sqrt{\frac{(l_2+1)(q_2-l_2)(p_2+q_2-l_2+1)}
                 {(k_2-l_2)(k_2-l_2+1)}} \nonumber \\
           &   & \times
                 \sqrt{\frac{(k_1-l_1+k_2-l_2-k+l)}
                            {(k-l+k_1-l_1-k_2+l_2+2)}}
                                                        \nonumber
  \label{eq:d}
  \end{eqnarray}
and
  \begin{equation}
  \nonumber
    \beta = \sqrt{\frac{(k+2)}{(k-q+1)(p+q-k)}}.
  \end{equation}

The $SU(3)$ CGC are found by using Equation \ref{eq:definitionisf}.
The corresponding $SU(2)$ coefficients are determined through Equation
\ref{eq:su2clebsch}.

\subsection{Algorithm for Isoscalar Factors}

The use of the above recursion relations to generate isoscalar factors
is far from trivial, and a detailed description of the algorithm is
given in \cite{williams3}.  A summary of that procedure is given here.

The ISFs for coupling to a state of highest weight can be thought of as
occupying lattice sites in an irregularly-shaped volume in a
three-dimensional space.  With the restriction provided by hypercharge
conservation, these ISFs can be thought of as functions of $s \equiv
I_1+I_2$, $k_1$, and $k_2$.  The two recursion relations above for the
shw states each have two terms with the same $s$ value, and two
others with $s$ values smaller by one $s-1$. All ISFs with the maximum
allowed $s$ value for a particular coupling are related, therefore, by
a two-term recursion relation.  Setting one of these equal to one (to
be fixed via normalization later), all others are simply constructed.
Because of the shaped of the lattice, one can in most cases find a
particular single ISF with $s=s_{\hbox{\scriptsize max}}-1$
which can be determined by
those with $s=s_{\hbox{\scriptsize max}}$:
once it is determined, all others with the
same $s$ value can be evaluated through the complete four-term recursion
relations based on values already known.  Repeating
this logic, one moves to a single ISF at the next lowest $s$ value from
which all others with the same $s$ can be determined, continuing until
all ISFs are known.  Normalization requires the sum of the squares of
all the ISFs so determined must be $1$, so each is multiplied by the
appropriate normalization factor.

When dealing with ISFs of degeneracy greater than one, it is necessary
to make several choices of ISF values in order for the algorithm to
succeed.  Which are chosen, and what values they are given, determines
the choice of degeneracy resolution.  The criteria mentioned in an
earlier section is equivalent to choosing ISFs for the $d-1$ highest
values of $s$ to be zero, and following the algorithm of the previous
paragraph beginning with the $s=s_{\hbox{\scriptsize max}}-d+1$ values.
A second set
begins with the $s=s_{\hbox{\scriptsize max}}-d+2$values,
and the further condition
of orthogonality of these ISFs to the previous ones.  Subsequent
sets of ISFs are formed likewise by successively starting with higher
values of $s$, and enforcing the condition of orthogonality to those
previously determined.

Regardless of degeneracy, there are particular combinations of
representations whose ISFs cannot be determined as above due to a
failure in the ability to step to lower values of $s$.  In every such
case, the algorithm succeeds for the conjugated ISFs,
and the exchange symmetry relation of Equation
\ref{defxi2} can be used to find the original values.

\section{Program Structure}

We define a variable type {\tt fraction} as an array of two long
integers, $n$ and $m$, to represent a number of the form $\sqrt{n/m}$.
All $SU(3)$ CGCs and ISFs can be exactly represented this way.
The routines that manipulate variables of the type {\tt fraction} are

\mbox{ }

\explanatory{reducefrac()}{Reduces a fraction to its lowest form.}
\explanatory{setequalfrac()}
            {Assigns one fraction to be equal to another.}
\explanatory{addfrac()}{Adds two fractions.}
\explanatory{subfrac()}{Subtracts two fractions.}
\explanatory{multfrac()}{Multiplies two fractions.}
\explanatory{imultfrac()}{Multiplies a fraction by a long integer.}
\explanatory{divfrac()}{Divides two fractions.}
\explanatory{absfrac()}{Changes a fraction to its absolute value and
                        returns its sign.}
\explanatory{factorialfrac()}
            {Finds the factorial of a fraction, if it is integral.}
\explanatory{equalfrac()}
            {Returns true if and only if two fractions are equal.}
\explanatory{comparefrac()}
            {Returns true if and only if the first fraction
                            is greater than the second.}
\explanatory{addirrat()}{Adds two numbers of the form $\sqrt{n/m}$.
                         Since the coefficients of $SU(3)$ always take
                         this form, no incompatibilities are expected.
                         An error message is written to the log file if
                         it is not possible to find
                       \begin{equation}
                       \nonumber
                         \sqrt{\frac{N}{M}} = \sqrt{\frac{n_1}{m_1}} +
                                              \sqrt{\frac{n_2}{m_2}}.
                       \end{equation}}

\mbox{ }

Global variables are

\mbox{ }

\explanatory{p1, q1, p2, q2}
            {The identities of the factor representations
                             $(p_1, q_1)$ and $(p_2, q_2)$.}
\explanatory{numprod}{The number of product representations in the
                      Clebsch-Gordan series.}
\explanatory{p[], q[]}{The identities of the product representations
                       $(p_i, q_i)$.}
\explanatory{names[]}
            {The names of the representations in the product.}
\explanatory{texnames[]}
            {The \LaTeX\ names of the representations in the product.}
\explanatory{ISFtable[][][][][][]}
            {Array which holds the isoscalar factors.
                       Its elements are of the type {\tt fraction}.}
\explanatory{SHWtable[][][][]}
            {Array which holds the isoscalar factors for the states of
                 highest weight.
                       Its elements are of the type {\tt fraction}.}
\explanatory{step\_down[][]}{Array which holds certain ratios of ISFs
                             for {\tt doshw()}.}
\explanatory{flag}{An integer flag used by {\tt doslice()} to control
                   {\tt doshw()}.}
\explanatory{errcount}
            {Cumulative count of errors in the program's execution.}
\explanatory{nameconv, labelconv}{Integers denoting the user's choices
                              for labelling conventions (see below).}
\explanatory{maxr, mink1, maxk1,
             maxl1, mink2, maxk2, maxl2}
            {Limits used in dimensioning the various arrays.}

\mbox{ }

Other subordinate routines included are

\mbox{ }

\explanatory{sizerep()}{Actually a macro which returns the dimension of
                        a representation.}
\explanatory{ipower()}{Raises a long integer to a power.}
\explanatory{factorial()}{Returns the factorial of a long integer.}
\explanatory{gcd()}{Returns the greatest common devisor of
                    two long integers.
                    It is employed by {\tt reducefrac()}.}
\explanatory{isqrt()}{Approximates the square root of a long integer
                      with another long integer.}
\explanatory{yii()}{Gives the hypercharge and isospin of a state with
                    given $k$, $l$, $m$.}
\explanatory{digitstring()}{Gives a character representing a one-digit
                            integer.  Used by {\tt numstring}.}
\explanatory{numstring()}{Returns a character string that represents an
                          integer.  Used by {\tt namerep()}.}
\explanatory{namerep()}{Gives the text name and
                        \LaTeX\ name of a representation.
                        Naming conventions are explained below.}
\explanatory{shw()}{Gives the hypercharge and isospin of the
                    highest-weight
                    state of a representation.}
\explanatory{initlims()}{Sets the values of {\tt mink1}, {\tt maxk1},
                         {\tt maxl1}, {\tt mink2}, {\tt maxk2},
                         {\tt maxl2},
                         and {\tt maxr}.}
\explanatory{inittable()}{Allocates memory space for {\tt ISFtable}
                          and initializes its values to zero.}
\explanatory{kill\_table()}{Releases the memory used by {\tt ISFtable}.}
\explanatory{putentry()}{Places an entry into {\tt ISFtable}.  Because
                         of hypercharge conservation, one index is
                         redundant for {\tt ISFtable};  this is used
                         by {\tt putentry()} and {\tt getentry()} to
                         reduce the size of the array
                         for {\tt ISFtable}.}
\explanatory{getentry()} {Checks to see if indices defining an
                          ISF are all in range, and if so, returns
                          an entry from {\tt ISFtable}.}
\explanatory{initshw()}{Allocates memory space for {\tt SHWtable}
                          and initializes its values to zero.}
\explanatory{killshw()}{Releases the memory used by {\tt SHWtable}.}
\explanatory{putentrySHW()}{Places an entry into {\tt SHWtable}.}
\explanatory{getentrySHW()}{Returns an entry from {\tt SHWtable}.}
\explanatory{printfraction()}{Writes a fraction in \LaTeX\ format to one
                              of the output files.}
\explanatory{printline()}{Writes a line to both {\tt cle.tex} and
                          {\tt iso.tex}.}
\explanatory{numstatesi()}{Counts the isomultiplets of given hypercharge
                         and isospin in the product.}
\explanatory{numstatesc()}{Counts the states of given hypercharge
                         and third component of isospin in the product.}
\explanatory{getconvs()}{Prompts for and
                         inputs the naming and labelling conventions
                         (see below) from the user.}

\mbox{ }

The central routines are

\mbox{ }

\explanatory{su2clebsch()}{Calculates an $SU(2)$ Clebsch-Gordan
                           coefficient
                           with Equation \ref{eq:su2clebsch}.}
\explanatory{degeneracy()}{Returns the number of times a representation
                           appears in the Clebsch-Gordan series.}
\explanatory{series()}{Finds the representations in the
                     Clebsch-Gordan series.
                     Also sets the value of {\tt numprod}.}
\explanatory{texphases()}{Tabulates the symmetry phases $\xi_1$ and
                          $\xi_3$
                          into {\tt iso.tex}.}
\explanatory{normhor()}{Checks that each row (for each isospin) of the
                        ISF table has unit norm.}
\explanatory{normvert()}{Checks that each column of the ISF table
                         has unit norm.}
\explanatory{checktable()}{Calls {\tt normhor()} and {\tt normvert()}.}
\explanatory{checksym()}{Checks that the shw ISFs are symmetric under
                         the exchange of {\bf r}$_1$ and {\bf r}$_2$,
                         in the cases where $p1=p2$, $q1=q2$.
                         It compares only the absolute values of
                         the ISFs.
                         It is called by {\tt doslice()}.}
\explanatory{textable()}{Loops over the quantum numbers of the factors
                         and products and tabulates the isoscalar
                         factors into {\tt iso.tex}.
                         Calls {\tt su2clebsch} and tabulates the
                         Clebsch-Gordan coefficients into
                         {\tt cle.tex}.}
\explanatory{coef{\it x}()}{Returns a coefficient of the recursion
                            relations
                            in Equations \ref{eq:a}-\ref{eq:d}.
                            Here {\it x} is one of
                            {\tt a1}, {\tt a2}, {\tt a3}, {\tt a4},
                            {\tt b1}, {\tt b2}, {\tt b3}, {\tt b4},
                            {\tt c1}, {\tt c2}, {\tt c3}, {\tt c4},
                            {\tt d1}, {\tt d2}, {\tt d3},
                            {\tt alpha}, {\tt beta}.}
\explanatory{l1step()}{Steps up or down in $l_1$ for {\tt doshw()}.}
\explanatory{k1step()}{Steps up or down in $k_1$ for {\tt doshw()}.}
\explanatory{doshw()}{Fills the parts of {\tt ISFtable} for the
                      highest-weight
                      states of the product representations.}
\explanatory{doslice()}{Fills {\tt ISFtable} for a given $(p_i, q_i)$
                        in the product.  Uses the algorithm given in
                        the previous section.  Calls {\tt doshw()}
                        to handle the states of highest weight.
                        Calls {\tt checksym()} for cases of
                        $p1=p2$, $q1=q2$.}
\explanatory{main()}{Prompts for $p_1$, $q_1$, $p_2$, $q_2$.
                     Calls {\tt getconvs()} to input labelling and
                     naming conventions.
                     Uses {\tt degeneracy} to determine the identities
                     of the
                     product representations and calls {\tt doslice()}
                     for each.
                     Calls {\tt textable()} to generate output.
                     Also calls {\tt normhor()} and {\tt normvert()}.}

\section{Using the Program}

\subsection{User Input}

The user, upon initiating the code, is queried to supply the identities
of the factor representations in the form $(p_1, q_1)$, $(p_2, q_2)$,
where each is an integer.  The naming and labelling conventions must
also be chosen.  The choices for naming convention are
  \begin{enumerate}
    \item  Representations are named as in \cite{slansky}, which is
           popular
           among particle physicists.  Representations are named by
           their dimension and bars are added based on triality
             \begin{equation}
             \nonumber
               t = (p - q) \mod 3.
             \end{equation}
           Representations with $t=1$ are unbarred, and those with $t=2$
           are barred.  One exception is that {\bf 6} $\equiv$ (2, 0).
           Representations with $t=0$ have bars if and only if $q > p$.
    \item  Representations are named by their dimension and bars are
           added if and only if $q > p$.
    \item  Representations are named by $(p, q)$.
  \end{enumerate}
In choices 1 and 2 above, there is an ambiguity when two representations
with different $(p, q)$ have the same dimension.  We distinguish them
using primes.  In the naming scheme 1 the lowest-lying
representations with same dimensions are
  \begin{equation}
  \nonumber
  \begin{array}{lclclclclcl}
    {\bf 15}  & = & (2, 1), & \quad &
    {\bf 15'} & = & (4, 0), \\
    {\bf 105} & = & (6, 2), & \quad &
    {\bf 105'} & = & (13, 0), \\
    {\bf 120} & = & (3, 5), & \quad &
    {\bf 120'} & = & (1, 9), & \quad &
    {\bf 120''} & = & (0, 14), \\
    {\bf 195} & = & (9, 2), & \quad &
    {\bf 195'} & = & (1, 12), \\
    {\bf 210} & = & (4, 6), & \quad &
    {\bf 210'} & = & (19, 0), \\
    {\bf 231} & = & (2, 10), & \quad &
    {\bf 231'} & = & (0, 20).
  \end{array}
  \end{equation}
The choices for labelling are
  \begin{enumerate}
    \item  States are labelled by hypercharge and isospin $y$, $i$,
           $i_z$.
           Isomultiplets are labelled by $y$ and $i$.
    \item  States are labelled by projection quantum numbers $k$,
           $l$, $m$.
           Isomultiplets are labelled by $k$ and $l$.
  \end{enumerate}

\subsection{Error Messages}

The program generates three files during its execution.
The log file ({\tt logfile} = {\tt su3.log}) contains messages and
reports the progress of the program.
See Sample 1 below.
Any error messages are written to the logfile.
The most important errors are also written to the console.
They are:

{\tt WARNING:  insufficient memory}\\
indicates that the free memory in the computer is too small to allow
dimensioning of the needed arrays.

{\tt WARNING:  integer overflow}\\
indicates that the integer mathematics has generated numbers larger
than the ``long'' integers defined by the computer.

{\tt WARNING:  Table is not horizontally normal}\\
indicates that the table of ISFs has nonnormal rows.
A row is given by $k1, l1, k2, l2$, and the isospin of the
states in the product.

{\tt WARNING:  Table is not vertically normal}\\
indicates that the table of ISFs has nonnormal columns.
A column is given by ${\bf R}, k, l$.

In addition, for cases in which $p1=p2$ and $q1=q2$, a quick check
is made of the exchange symmetry.
If a nonsymmetric entry is found, the message

{\tt WARNING:  n=4 SHW ISFs are not symmetric}

\noindent
(for example) appears.

These errors indicate that a serious problem has occurred.
Other error messages are written only to {\tt logfile} and are
self-explanatory.

\subsection{Output}

The ISFs are written in \LaTeX\ format to the file
{\tt isofile} = {\tt iso.tex}.
In it, representations are named according to the naming choice made at
the beginning of execution, and states are labelled according to the
labelling choice.
Multiply degenerate representations are distinguished by subscripts.
A square root is assumed to appear over the unsigned part of each entry.
Thus,
  \begin{equation}
  \nonumber
  \begin{array}{rrrr|c|}
    &&&& {\bf R} \\
    \multicolumn{4}{c|}{{\bf r}_1 \; {\otimes} \; {\bf r}_2} & Y \\
    &&&& I \\
  \hline
    y_1 & i_1 & y_2 & i_2 & \pm C \\
  \hline
  \end{array}
  \end{equation}
means that the isoscalar factor
  \begin{equation}
  \nonumber
    F({\bf R}, Y, I: {\bf r}_1, y_1, i_1; {\bf r}_2, y_2, i_2)
        = \pm \sqrt{C}
  \end{equation}
in the ($y$, $i$) labelling convention.
In the ($k$, $l$) convention, this would appear as
  \begin{equation}
  \nonumber
  \begin{array}{rrrr|c|}
    &&&& {\bf R} \\
    \multicolumn{4}{c|}{{\bf r}_1 \; {\otimes} \; {\bf r}_2} & k \\
    &&&& l \\
  \hline
    k_1 & l_1 & k_2 & l_2 & \pm C \\
  \hline
  \end{array}
  \end{equation}
The symmetry factors $\xi_1$ and $\xi_2$ are also given in this file.
For example,
  \begin{equation}
  \nonumber
  \begin{array}{c|c|}
    {\bf r}_1 \; \otimes \; {\bf r}_2 & {\bf R} \\
  \hline
    \xi_1 & + \\
    \xi_2 & - \\
  \hline
  \end{array}
  \end{equation}
means that
  \begin{equation}
  \nonumber
  \begin{array}{lcl}
    \xi_1 ({\bf R}: {\bf r}_1; {\bf r}_2) &=& +1, \\
    \xi_2 ({\bf R}: {\bf r}_1; {\bf r}_2) &=& -1.
  \end{array}
  \end{equation}

The Clebsch-Gordan coefficients are written in \LaTeX\ format to the
file {\tt clefile} = {\tt cle.tex}.
Again, a square root is assumed over the unsigned part of each entry.
In the $(y, i, i_z)$ labelling convention, the format is
  \begin{equation}
  \nonumber
  \begin{array}{rrrrrr|c|}
    &&&&&& {\bf R} \\
    &&&&&& Y \\
    \multicolumn{6}{c|}{{\bf r}_1 \; {\otimes} \; {\bf r}_2}
         & I \\
    &&&&&& I_z \\
  \hline
    y_1 & i_1 & i_{1z} & y_2 & i_2 & i_{2z} & \pm C \\
  \hline
  \end{array}
  \end{equation}
for
  \begin{equation}
  \nonumber
    \left< {\bf R} \; Y \; I\; I_z| {\bf r}_1 \; y_1 \; i_1 \; i_{1z}
                         \; {\bf r}_2 \; y_2 \; i_2 \; i_{2z} \right>
        = \pm \sqrt{C}.
  \end{equation}
In the $(k, l, m)$ notation this is
  \begin{equation}
  \nonumber
  \begin{array}{rrrrrr|c|}
    &&&&&& {\bf R} \\
    &&&&&& k \\
    \multicolumn{6}{c|}{{\bf r}_1 \; {\otimes} \; {\bf r}_2}
         & l \\
    &&&&&& m \\
  \hline
    k_1 & l_1 & m_1 & k_2 & l_2 & m_2 & \pm C \\
  \hline
  \end{array}
  \end{equation}

Samples 2 and 3 below show examples of {\tt iso.tex} and {\tt cle.tex}
after processing by \LaTeX.

\section{Sample Output}

Presented here are sample output files from a successful run.
The product computed is {\bf 15} $\otimes$ {\bf 8} = (2, 1) $\otimes$
(1, 1), which has a doubly degenerate {\bf 15}.  In this run, we have
chosen the naming convention of \cite{slansky} and the labelling
convention of hypercharge and isospin ($y$, $i$).
Sample 1 is the log file {\tt su3.log}.
It echoes our naming and labelling conventions, informs us of the
program's progress, and tells us that the table of isoscalar factors
is properly normalized.
Sample 2 is the \LaTeX\ file containing the isoscalar factors.
The Clebsch-Gordan coefficients are in the \LaTeX\ file of Sample 3.
Notation for the tables is explained in the previous section.

\section{Acknowledgement}

The subroutine {\tt gcd()} was coded by Tom Whaley.

Part of this work (TAK)
was supported by the Director, Office of Energy
Research, Office of High Energy and Nuclear Physics, Division of High
Energy Physics of the U.S. Department of Energy under Contract
DE-AC03-76SF00098.

\newpage

\begin{center}
{\large \bf SAMPLE CAPTIONS}
\end{center}
\vskip .25in

\samplecaption{1}{Log file ({\tt su3.log}) from a sample run calculating
                  {\bf 15} $\otimes$ {\bf 8} = $(2,1) \otimes (1,1)$..}

\samplecaption{2}{Output file ({\tt iso.tex}) from a run calculating
                  {\bf 15} $\otimes$ {\bf 8}.
                  These are the isoscalar factors.
                  Notation is explained in Section 4.
                  The output has already been processed by \LaTeX.
                  Only the first page is shown, in order to save space.}

\samplecaption{3}{Output file ({\tt cle.tex}) from a run calculating
                  {\bf 15} $\otimes$ {\bf 8}.
                  These are the Clebsch-Gordan coefficients.
                  Notation is explained in Section 4.
                  The output has already been processed by \LaTeX.
                  Only the first page of the file is shown, in order
                  to save space.}

\newpage
\renewcommand \pagestyle{empty}

\begin{flushleft}
\begin{tt}
SU3:~~beginning\\
{}~\\
getconvs:~~naming convention is of Slansky\\
getconvs:~~labelling convention is (y, i, i3)\\
{}~\\
SU3:~~calculating 15 x 8\\
{}~\\
SU3:~~working on 3 with degeneracy = 1\\
SU3:~~working on 6* with degeneracy = 1\\
SU3:~~working on 15' with degeneracy = 1\\
SU3:~~working on 15 with degeneracy = 2\\
SU3:~~working on 24 with degeneracy = 1\\
SU3:~~working on 42 with degeneracy = 1\\
{}~\\
checktable:~~table is horizontally normal\\
checktable:~~table is vertically normal\\
{}~\\
SU3:~~writing output files\\
{}~\\
SU3:~~ending\\
\end{tt}
\end{flushleft}

\newpage

\leftmargin 0in
\textwidth 6in
\textheight 9in

\begin{center}

Isoscalar Factors for {\bf 15} $\otimes$ {\bf 8} \\

\space

\begin{tabular}{c|ccccccc|}
{\bf 15} $\otimes$ {\bf 8}
 & {\bf 3}
 & $\bar{\bf 6}$
 & {\bf 15}${\bf '}$
 & {\bf 15}$_1$
 & {\bf 15}$_2$
 & {\bf 24}
 & {\bf 42}
\\
\hline
$\xi_1$
 & $+$
 & $+$
 & $-$
 & $-$
 & $+$
 & $-$
 & $+$
\\
$\xi_3$
 & $+$
 & $-$
 & $+$
 & $-$
 & $+$
 & $+$
 & $+$
\\
\hline
\end{tabular}

\space

\end{center}

\begin{raggedright}

\begin{tabular}{rrrr|r|}
 & & &
 & {\bf 24}
\\
 & & &
 & \f{7}{3}
\\
 & & &
 & \f{1}{2}
\\
\hline
   \f{4}{3}
 & 1
 & 1
 & \f{1}{2}
    & -1
 \\
\hline
\end{tabular}
\mbox{ }
\begin{tabular}{rrrr|r|}
 & & &
 & {\bf 42}
\\
 & & &
 & \f{7}{3}
\\
 & & &
 & \f{3}{2}
\\
\hline
   \f{4}{3}
 & 1
 & 1
 & \f{1}{2}
    & 1
 \\
\hline
\end{tabular}
\mbox{ }
\begin{tabular}{rrrr|rr|}
 & & &
 & $\bar{\bf 6}$
 & {\bf 24}
\\
 & & &
 & \f{4}{3}
 & \f{4}{3}
\\
 & & &
 & 0
 & 0
\\
\hline
   \f{1}{3}
 & \f{1}{2}
 & 1
 & \f{1}{2}
    & \f{1}{5}
    & -\f{4}{5}
 \\
   \f{4}{3}
 & 1
 & 0
 & 1
    & -\f{4}{5}
    & -\f{1}{5}
 \\
\hline
\end{tabular}
\mbox{ }
\begin{tabular}{rrrr|rrrr|}
 & & &
 & {\bf 15}$_1$
 & {\bf 15}$_2$
 & {\bf 24}
 & {\bf 42}
\\
 & & &
 & \f{4}{3}
 & \f{4}{3}
 & \f{4}{3}
 & \f{4}{3}
\\
 & & &
 & 1
 & 1
 & 1
 & 1
\\
\hline
   \f{1}{3}
 & \f{1}{2}
 & 1
 & \f{1}{2}
    & -\f{400}{1647}
    & -\f{1}{61}
    & -\f{4}{27}
    & \f{16}{27}
 \\
   \f{1}{3}
 & \f{3}{2}
 & 1
 & \f{1}{2}
    & \f{529}{3294}
    & -\f{32}{61}
    & -\f{8}{27}
    & -\f{1}{54}
 \\
   \f{4}{3}
 & 1
 & 0
 & 0
    & \f{49}{183}
    & -\f{4}{61}
    & \f{1}{3}
    & \f{1}{3}
 \\
   \f{4}{3}
 & 1
 & 0
 & 1
    & \f{361}{1098}
    & \f{24}{61}
    & -\f{2}{9}
    & \f{1}{18}
 \\
\hline
\end{tabular}
\mbox{ }
\begin{tabular}{rrrr|rr|}
 & & &
 & {\bf 15}${\bf '}$
 & {\bf 42}
\\
 & & &
 & \f{4}{3}
 & \f{4}{3}
\\
 & & &
 & 2
 & 2
\\
\hline
   \f{1}{3}
 & \f{3}{2}
 & 1
 & \f{1}{2}
    & \f{1}{2}
    & \f{1}{2}
 \\
   \f{4}{3}
 & 1
 & 0
 & 1
    & -\f{1}{2}
    & \f{1}{2}
 \\
\hline
\end{tabular}
\mbox{ }
\begin{tabular}{rrrr|rrrrrr|}
 & & &
 & {\bf 3}
 & $\bar{\bf 6}$
 & {\bf 15}$_1$
 & {\bf 15}$_2$
 & {\bf 24}
 & {\bf 42}
\\
 & & &
 & \f{1}{3}
 & \f{1}{3}
 & \f{1}{3}
 & \f{1}{3}
 & \f{1}{3}
 & \f{1}{3}
\\
 & & &
 & \f{1}{2}
 & \f{1}{2}
 & \f{1}{2}
 & \f{1}{2}
 & \f{1}{2}
 & \f{1}{2}
\\
\hline
   -\f{2}{3}
 & 0
 & 1
 & \f{1}{2}
    & \f{1}{40}
    & \f{3}{20}
    & -\f{50}{183}
    & -\f{9}{488}
    & -\f{4}{15}
    & \f{4}{15}
 \\
   -\f{2}{3}
 & 1
 & 1
 & \f{1}{2}
    & -\f{3}{20}
    & \f{1}{10}
    & \f{121}{549}
    & -\f{75}{244}
    & -\f{8}{45}
    & -\f{2}{45}
 \\
   \f{1}{3}
 & \f{1}{2}
 & 0
 & 0
    & -\f{9}{80}
    & -\f{3}{40}
    & \f{4}{183}
    & -\f{121}{976}
    & \f{2}{15}
    & \f{8}{15}
 \\
   \f{1}{3}
 & \f{1}{2}
 & 0
 & 1
    & -\f{1}{80}
    & -\f{5}{24}
    & \f{196}{1647}
    & \f{225}{976}
    & -\f{10}{27}
    & \f{8}{135}
 \\
   \f{1}{3}
 & \f{3}{2}
 & 0
 & 1
    & \f{2}{5}
    & -\f{4}{15}
    & \f{2}{1647}
    & -\f{18}{61}
    & -\f{4}{135}
    & -\f{1}{135}
 \\
   \f{4}{3}
 & 1
 & -1
 & \f{1}{2}
    & \f{3}{10}
    & \f{1}{5}
    & \f{200}{549}
    & \f{3}{122}
    & \f{1}{45}
    & \f{4}{45}
 \\
\hline
\end{tabular}
\mbox{ }
\begin{tabular}{rrrr|rrrrr|}
 & & &
 & {\bf 15}${\bf '}$
 & {\bf 15}$_1$
 & {\bf 15}$_2$
 & {\bf 24}
 & {\bf 42}
\\
 & & &
 & \f{1}{3}
 & \f{1}{3}
 & \f{1}{3}
 & \f{1}{3}
 & \f{1}{3}
\\
 & & &
 & \f{3}{2}
 & \f{3}{2}
 & \f{3}{2}
 & \f{3}{2}
 & \f{3}{2}
\\
\hline
   -\f{2}{3}
 & 1
 & 1
 & \f{1}{2}
    & \f{1}{4}
    & -\f{361}{2196}
    & -\f{12}{61}
    & -\f{1}{9}
    & \f{5}{18}
 \\
   \f{1}{3}
 & \f{1}{2}
 & 0
 & 1
    & -\f{1}{3}
    & -\f{1}{1647}
    & \f{9}{61}
    & -\f{4}{27}
    & \f{10}{27}
 \\
   \f{1}{3}
 & \f{3}{2}
 & 0
 & 0
    & \f{3}{16}
    & \f{25}{2928}
    & \f{16}{61}
    & \f{1}{3}
    & \f{5}{24}
 \\
   \f{1}{3}
 & \f{3}{2}
 & 0
 & 1
    & \f{5}{48}
    & \f{305}{432}
    & 0
    & -\f{5}{27}
    & \f{1}{216}
 \\
   \f{4}{3}
 & 1
 & -1
 & \f{1}{2}
    & -\f{1}{8}
    & \f{529}{4392}
    & -\f{24}{61}
    & \f{2}{9}
    & \f{5}{36}
 \\
\hline
\end{tabular}
\mbox{ }
\begin{tabular}{rrrr|r|}
 & & &
 & {\bf 42}
\\
 & & &
 & \f{1}{3}
\\
 & & &
 & \f{5}{2}
\\
\hline
   \f{1}{3}
 & \f{3}{2}
 & 0
 & 1
    & 1
 \\
\hline
\end{tabular}
\mbox{ }
\begin{tabular}{rrrr|rrrr|}
 & & &
 & {\bf 3}
 & {\bf 15}$_1$
 & {\bf 15}$_2$
 & {\bf 42}
\\
 & & &
 & -\f{2}{3}
 & -\f{2}{3}
 & -\f{2}{3}
 & -\f{2}{3}
\\
 & & &
 & 0
 & 0
 & 0
 & 0
\\
\hline
   -\f{5}{3}
 & \f{1}{2}
 & 1
 & \f{1}{2}
    & -\f{2}{5}
    & \f{49}{122}
    & -\f{6}{61}
    & -\f{1}{10}
 \\
   -\f{2}{3}
 & 0
 & 0
 & 0
    & -\f{3}{20}
    & -\f{3}{61}
    & -\f{49}{244}
    & \f{3}{5}
 \\
   -\f{2}{3}
 & 1
 & 0
 & 1
    & \f{3}{10}
    & \f{1}{366}
    & -\f{81}{122}
    & -\f{1}{30}
 \\
   \f{1}{3}
 & \f{1}{2}
 & -1
 & \f{1}{2}
    & \f{3}{20}
    & \f{100}{183}
    & \f{9}{244}
    & \f{4}{15}
 \\
\hline
\end{tabular}
\mbox{ }

\end{raggedright}

\newpage

\topmargin -.5in

\begin{center}

Clebsch-Gordan Coefficients for {\bf 15} $\otimes$ {\bf 8} \\

\space

\end{center}

\begin{raggedright}

\begin{tabular}{rrrrrr|r|}
 & & & & &
 & {\bf 42}
\\
 & & & & &
 & \f{7}{3}
\\
 & & & & &
 & \f{3}{2}
\\
 & & & & &
 & -\f{3}{2}
\\
\hline
   \f{4}{3}
 & 1
 & -1
 & 1
 & \f{1}{2}
 & -\f{1}{2}
    & 1
 \\
\hline
\end{tabular}
\mbox{ }
\begin{tabular}{rrrrrr|rr|}
 & & & & &
 & {\bf 24}
 & {\bf 42}
\\
 & & & & &
 & \f{7}{3}
 & \f{7}{3}
\\
 & & & & &
 & \f{1}{2}
 & \f{3}{2}
\\
 & & & & &
 & -\f{1}{2}
 & -\f{1}{2}
\\
\hline
   \f{4}{3}
 & 1
 & -1
 & 1
 & \f{1}{2}
 & \f{1}{2}
    & \f{2}{3}
    & \f{1}{3}
 \\
   \f{4}{3}
 & 1
 & 0
 & 1
 & \f{1}{2}
 & -\f{1}{2}
    & -\f{1}{3}
    & \f{2}{3}
 \\
\hline
\end{tabular}
\mbox{ }
\begin{tabular}{rrrrrr|rr|}
 & & & & &
 & {\bf 24}
 & {\bf 42}
\\
 & & & & &
 & \f{7}{3}
 & \f{7}{3}
\\
 & & & & &
 & \f{1}{2}
 & \f{3}{2}
\\
 & & & & &
 & \f{1}{2}
 & \f{1}{2}
\\
\hline
   \f{4}{3}
 & 1
 & 0
 & 1
 & \f{1}{2}
 & \f{1}{2}
    & \f{1}{3}
    & \f{2}{3}
 \\
   \f{4}{3}
 & 1
 & 1
 & 1
 & \f{1}{2}
 & -\f{1}{2}
    & -\f{2}{3}
    & \f{1}{3}
 \\
\hline
\end{tabular}
\mbox{ }
\begin{tabular}{rrrrrr|r|}
 & & & & &
 & {\bf 42}
\\
 & & & & &
 & \f{7}{3}
\\
 & & & & &
 & \f{3}{2}
\\
 & & & & &
 & \f{3}{2}
\\
\hline
   \f{4}{3}
 & 1
 & 1
 & 1
 & \f{1}{2}
 & \f{1}{2}
    & 1
 \\
\hline
\end{tabular}
\mbox{ }

\begin{tabular}{rrrrrr|rr|}
 & & & & &
 & {\bf 15}${\bf '}$
 & {\bf 42}
\\
 & & & & &
 & \f{4}{3}
 & \f{4}{3}
\\
 & & & & &
 & 2
 & 2
\\
 & & & & &
 & -2
 & -2
\\
\hline
   \f{1}{3}
 & \f{3}{2}
 & -\f{3}{2}
 & 1
 & \f{1}{2}
 & -\f{1}{2}
    & \f{1}{2}
    & \f{1}{2}
 \\
   \f{4}{3}
 & 1
 & -1
 & 0
 & 1
 & -1
    & -\f{1}{2}
    & \f{1}{2}
 \\
\hline
\end{tabular}
\mbox{ }
\begin{tabular}{rrrrrr|rrrrrr|}
 & & & & &
 & {\bf 15}${\bf '}$
 & {\bf 15}$_1$
 & {\bf 15}$_2$
 & {\bf 24}
 & {\bf 42}
 & {\bf 42}
\\
 & & & & &
 & \f{4}{3}
 & \f{4}{3}
 & \f{4}{3}
 & \f{4}{3}
 & \f{4}{3}
 & \f{4}{3}
\\
 & & & & &
 & 2
 & 1
 & 1
 & 1
 & 1
 & 2
\\
 & & & & &
 & -1
 & -1
 & -1
 & -1
 & -1
 & -1
\\
\hline
   \f{1}{3}
 & \f{1}{2}
 & -\f{1}{2}
 & 1
 & \f{1}{2}
 & -\f{1}{2}
    & 0
    & -\f{400}{1647}
    & -\f{1}{61}
    & -\f{4}{27}
    & \f{16}{27}
    & 0
 \\
   \f{1}{3}
 & \f{3}{2}
 & -\f{3}{2}
 & 1
 & \f{1}{2}
 & \f{1}{2}
    & \f{1}{8}
    & -\f{529}{4392}
    & \f{24}{61}
    & \f{2}{9}
    & \f{1}{72}
    & \f{1}{8}
 \\
   \f{1}{3}
 & \f{3}{2}
 & -\f{1}{2}
 & 1
 & \f{1}{2}
 & -\f{1}{2}
    & \f{3}{8}
    & \f{529}{13176}
    & -\f{8}{61}
    & -\f{2}{27}
    & -\f{1}{216}
    & \f{3}{8}
 \\
   \f{4}{3}
 & 1
 & -1
 & 0
 & 0
 & 0
    & 0
    & \f{49}{183}
    & -\f{4}{61}
    & \f{1}{3}
    & \f{1}{3}
    & 0
 \\
   \f{4}{3}
 & 1
 & -1
 & 0
 & 1
 & 0
    & -\f{1}{4}
    & -\f{361}{2196}
    & -\f{12}{61}
    & \f{1}{9}
    & -\f{1}{36}
    & \f{1}{4}
 \\
   \f{4}{3}
 & 1
 & 0
 & 0
 & 1
 & -1
    & -\f{1}{4}
    & \f{361}{2196}
    & \f{12}{61}
    & -\f{1}{9}
    & \f{1}{36}
    & \f{1}{4}
 \\
\hline
\end{tabular}
\mbox{ }
\begin{tabular}{rrrrrr|rrrrrrrr|}
 & & & & &
 & $\bar{\bf 6}$
 & {\bf 15}${\bf '}$
 & {\bf 15}$_1$
 & {\bf 15}$_2$
 & {\bf 24}
 & {\bf 24}
 & {\bf 42}
 & {\bf 42}
\\
 & & & & &
 & \f{4}{3}
 & \f{4}{3}
 & \f{4}{3}
 & \f{4}{3}
 & \f{4}{3}
 & \f{4}{3}
 & \f{4}{3}
 & \f{4}{3}
\\
 & & & & &
 & 0
 & 2
 & 1
 & 1
 & 0
 & 1
 & 1
 & 2
\\
 & & & & &
 & 0
 & 0
 & 0
 & 0
 & 0
 & 0
 & 0
 & 0
\\
\hline
   \f{1}{3}
 & \f{1}{2}
 & -\f{1}{2}
 & 1
 & \f{1}{2}
 & \f{1}{2}
    & -\f{1}{10}
    & 0
    & -\f{200}{1647}
    & -\f{1}{122}
    & \f{2}{5}
    & -\f{2}{27}
    & \f{8}{27}
    & 0
 \\
   \f{1}{3}
 & \f{1}{2}
 & \f{1}{2}
 & 1
 & \f{1}{2}
 & -\f{1}{2}
    & \f{1}{10}
    & 0
    & -\f{200}{1647}
    & -\f{1}{122}
    & -\f{2}{5}
    & -\f{2}{27}
    & \f{8}{27}
    & 0
 \\
   \f{1}{3}
 & \f{3}{2}
 & -\f{1}{2}
 & 1
 & \f{1}{2}
 & \f{1}{2}
    & 0
    & \f{1}{4}
    & -\f{529}{6588}
    & \f{16}{61}
    & 0
    & \f{4}{27}
    & \f{1}{108}
    & \f{1}{4}
 \\
   \f{1}{3}
 & \f{3}{2}
 & \f{1}{2}
 & 1
 & \f{1}{2}
 & -\f{1}{2}
    & 0
    & \f{1}{4}
    & \f{529}{6588}
    & -\f{16}{61}
    & 0
    & -\f{4}{27}
    & -\f{1}{108}
    & \f{1}{4}
 \\
   \f{4}{3}
 & 1
 & -1
 & 0
 & 1
 & 1
    & -\f{4}{15}
    & -\f{1}{12}
    & -\f{361}{2196}
    & -\f{12}{61}
    & -\f{1}{15}
    & \f{1}{9}
    & -\f{1}{36}
    & \f{1}{12}
 \\
   \f{4}{3}
 & 1
 & 0
 & 0
 & 0
 & 0
    & 0
    & 0
    & \f{49}{183}
    & -\f{4}{61}
    & 0
    & \f{1}{3}
    & \f{1}{3}
    & 0
 \\
   \f{4}{3}
 & 1
 & 0
 & 0
 & 1
 & 0
    & \f{4}{15}
    & -\f{1}{3}
    & 0
    & 0
    & \f{1}{15}
    & 0
    & 0
    & \f{1}{3}
 \\
   \f{4}{3}
 & 1
 & 1
 & 0
 & 1
 & -1
    & -\f{4}{15}
    & -\f{1}{12}
    & \f{361}{2196}
    & \f{12}{61}
    & -\f{1}{15}
    & -\f{1}{9}
    & \f{1}{36}
    & \f{1}{12}
 \\
\hline
\end{tabular}
\mbox{ }

\end{raggedright}


\newpage

\begin{thebibliography}{99}

\bibitem{deswart}      J.\ J.\ de Swart, \rmp{35}{916}{63},
                           \rmp{37}{326}{65}.
\bibitem{carruthers}   P.\ A.\ Carruthers, Introduction to Unitary
                           Symmetry
                           (John Wiley \& Sons, New York, 1966).
\bibitem{georgi}       H.\ Georgi, Lie Algebras in Particle Physics
                           (Benjamin/Cummings, London, 1982).
\bibitem{bied}         G.\ E.\ Baird and L.\ C.\ Biedenharn,
                           \jmp{4}{436}{63},
                           \blank{4}{1449}{63}, \blank{5}{1723}{64},
                           \blank{5}{1730}{64}.
\bibitem{chilton}      P.\ McNamee and F.\ Chilton, \rmp{36}{1005}{64}.
\bibitem{wali}         V.\ Rabl, G.\ Campbell, Jr., and K.\ C.\ Wali,
                           \jmp{16}{2494}{75}.
\bibitem{hecht}        K.\ T.\ Hecht, \np{62}{1}{65}.
\bibitem{rashid}       M.\ A.\ Rashid, Nuovo Cim.\ \blank{26}{118}{62}.
\bibitem{kaeding2}     T.\ A.\ Kaeding,
                           Tables of $SU(3)$ Isoscalar Factors,
                           LBL-36659, UCB-PTH-95/01,
                           to appear in
                           Atomic Data and Nuclear Data Tables 61.
\bibitem{akiyama}      Y.\ Akiyama and J.\ P.\ Draayer,
                           \cpc{5}{405}{73}.
\bibitem{williams1}    H.\ T.\ Williams, Symmetry Properties of SU3
                           Vector Coupling Coefficients, WLUPY-9-93.
\bibitem{williams2}    H.\ T.\ Williams and C.\ J.\ Wynne,
                           Comp.\ in Phys.\ \blank{8}{355}{94}.
\bibitem{kaeding1}     T.\ A.\ Kaeding, \cpc{85}{82}{95}.
\bibitem{latexref}     L.\ Lamport, \LaTeX:
                           A Document Preparation System
                           (Addison-Wesley, Menlo Park CA, 1986).
\bibitem{oreilly}      M.\ F.\ O'Reilly, \jmp{23}{2022}{82}.
\bibitem{wigner2}      E.\ P.\ Wigner, Group Theory and its Application
                           to the Quantum Mechanics of Atomic Spectra,
                           trans.\ by J.\ J.\ Griffin
                           (Academic Press, New York, 1959).
\bibitem{condon}       E.\ U.\ Condon and G.\ H.\ Shortley, The Theory
                           of Atomic Spectra
                           (Cambridge University Press, London, 1957).
\bibitem{alcaras}      L.\ C.\ Biedenharn, \pl{3}{254}{63};
                       L.\ C.\ Biedenharn, A.\ Giovanni,
                           and L.\ D.\ Louck,
                           \jmp{8}{691}{67};
                       J.\ A.\ Alcaras, L.\ C.\ Biedenharn,
                           K.\ T.\ Hecht, and G.\ Neely,
                           Ann. Phys. \blank{60}{85}{70};
                       Z.\ Pluhar, Yu.\ F.\ Smirnov, V.\ N.\ Tolstoy,
                           \jp{A19}{21}{86};
\bibitem{bied2}        L.\ C.\ Biedenharn, M.\ A.\ Lohe,
                           and H.\ T.\ Williams, \jmp{35}{6672}{94}.
\bibitem{williams3}    H.\ T.\ Williams, SU3 Isoscalar Factors,
                           WLUPY995.1, hep-th/9509167,
                           submitted to Journal of Mathematical Physics.
\bibitem{slansky}      R.\ Slansky, \prep{79}{1}{81}.


\end{thebibliography}
\end{document}